\UseRawInputEncoding
\documentclass[preprint,dvips,numbers,sort&compress]{elsarticle}
\usepackage{hyperref}
\usepackage{amsmath}
\usepackage{amssymb}
\usepackage{amsfonts}
\usepackage{amsthm}
\usepackage{arydshln}
\usepackage[toc, page]{appendix}
\usepackage{vector}
\usepackage{graphicx}
\usepackage{subfigure}
\usepackage{color}
\usepackage{paralist}
\usepackage{multirow}
\usepackage{booktabs}
\usepackage{setspace}
\graphicspath{{shang_eps/}}

\numberwithin{equation}{section}
\begin{document}
\title{ 
On Polymer Statistical Mechanics: From Gaussian Distribution to Maxwell-Boltzmann Distribution to Fermi-Dirac Distribution}
\author[a1]{Lixiang Yang}
\address[a1]{Department of Mechanical and Aerospace Engineering, The Ohio State University, Columbus OH, 43210 Email: yang.1130@buckeyemail.osu.edu}
%

\begin{abstract} \noindent
Macroscopic mechanical properties of polymers are determined by their microscopic molecular chain distribution. Due to randomness of these molecular chains, probability theory has been used to find their micro-states and energy distribution. In this paper, aided by central limit theorem and mixed Bayes rule,
we showed that entropy elasticity based on Gaussian distribution is questionable. By releasing freely jointed chain assumption, we found that there is energy redistribution when each bond of a molecular chain changes its length. Therefore, we have to change Gaussian distribution used in polymer elasticity to Maxwell-Boltzmann distribution. Since Maxwell-Boltzmann distribution is only a good energy description for gas molecules, we found a mathematical path to change Maxwell-Boltzmann distribution to Fermi-Dirac distribution based on molecular chain structures. Because a molecular chain can be viewed as many monomers glued by covalent electrons, Fermi-Dirac distribution describes the probability of covalent electron occupancy in micro-states for solids such as polymers. Mathematical form of Fermi-Dirac distribution is logistic function. Mathematical simplicity of Fermi-Dirac distribution makes many hard mechanics problems easy to understand. Generalized logistic function or Fermi-Dirac distribution function was able to understand many polymer mechanics problems such as viscoelasticity \citep{yang_viscoelasticity_2021}, viscoplasticity \citep{yang_mathematical_2019}, shear band and necking \citep{yang_revisit_2020}, and ultrasonic bonding \citep{yang_digest_2023}. 

\hskip 10pt
\noindent {\bf Keywords: Gaussian Statistical Mechanics; Polymer Physics; Mixed Bayes Rule; Conditional Probability Density Functions; Maxwell-Boltzmann Distribution; Fermi-Dirac Distribution } 
\end{abstract}
\maketitle
\section{Introduction to Entropy Elasticity}
\label{s:intro} \noindent
Mechanical behaviors of rubbers are largely determined by its micro-structures \citep{zhan_new_2023,li_effects_2021, wang_mechanics_2015, akagi_examination_2011}. Since rubbers are microscopically made of randomly distributed molecular chains, probability theory is introduced to understand their stress-strain relationship \citep{buche_chain_2021}. When rubbers are stretched under external forces, molecular chains will get stretched and change their configurations. After a molecular chain changes from one configuration to another one, its entropy will change too. 

In entropy rubber elasticity, entropy instead of internal energy is considered as the driving force of mechanical response \citep{zhang_two-component_2022}. So force and displacement or stress-strain relationship is derived by relationship between probability distribution of molecular chains  and entropy of the system \citep{zhang_single-chain_2022}.

For a single chain, with one end fixed, the probability to find the other end of the chain in a small box will obey Gaussian distribution or normal distribution. Flory\citep{flory_principles_1953} derived Gaussian distribution for a random freely jointed chain in one dimension. The distance of a molecular chain from one end to the other end is projected to x-axis which can be treated as a random variable $S_n$. The probability distribution of $S_n$ can be considered as the corresponding excess of heads in a series of coin tosses. In another word, random variable $S_n$ is a binomial random variable with probability of each trial 0.5. In chapter X appendix A \citep{flory_principles_1953}, Flory also showed the random variable $S_n$ will become normal distribution if let the number of the bond vectors of a molecular chain go to infinity. One dimensional Gaussian distribution density function of end to end distance of a molecular chain in $x$ direction was given as
\begin{equation} 
G(x, N) = (\,\frac{3}{2 \pi Nb^2} )\,^{1/2}\mathbf{e}^{-3x^2/2Nb^2},   
\label{eq:Gauss_1D} 
\end{equation}
where $N$ is the number of Kuhn segments of a molecular chain and $b$ is given as Kuhn length.
By assuming Gaussian distributions in $x$, $y$ and $z$ directions are independent of each other, one dimensional Gaussian distribution can be extended to become three dimensional Gaussian distribution \citep{darabi_amended_2023}
\begin{equation} 
G(\mathbf{r}, N) = (\,\frac{3}{2 \pi Nb^2} )\,^{3/2}\mathbf{e}^{-3r^2/2Nb^2},   
\label{eq:Gauss_3D} 
\end{equation} 
where $r$ is the magnitude of the chain displacement vector $\mathbf{r}$, i.e.
\begin{equation*} 
r^2 = x^2 + y^2 + z^2.   
\label{eq:Gauss_3D_a} 
\end{equation*}

In fact, what Flory derived can be mathematically proved by using central limit theorem. Based on central limit theorem, the number of segment of each molecular chain is not necessarily going to infinity to be a Gaussian distribution. In many cases, Gaussian distribution can be approached even with several segments if probability distribution function of each segment is symmetric.

Once probability of end to end distance of a molecular chain is obtained, entropy of this single chain is related to its micro-states by Boltzmann's entropy equation. 
If all molecular chains are taken to deform affinely with the macroscopic deformation,  probability distribution of a single molecular chain can be extended to an entire polymer system. For an entire polymer system,  Boltzmann's entropy equation can be written as
\begin{equation} 
S \: = \: k_B ln \Omega,  
\label{eq:boltzmann} 
\end{equation}
where $S$ is entropy of an entire polymer system and $k_B = 1.3807 \times 10^{-23} JK^{-1}$ is Boltzmann's constant. $\Omega$ is the number of micro-states that this system can have.  Originally developed for idea gas, Boltzmann's entropy equation was firstly introduced by Werner Kuhn in 1930s to model rubber elasticity. This method was later adopted by Paul Flory and many other researchers \citep{tanaka_viscoelastic_1992, vernerey_statistical_2018,mahnken_statistically_2022, jiang_strain_2022}. This entropy concept still dominates rubber elasticity up to nowadays.
 Boltzmann's entropy equation can be derived by using a statistical representation of temperature and the first law of thermodynamics.  
If a system goes to equilibrium, it will select a macroscopic configuration that maximizes the number of micro-states.  If we assume that there is only one single micro-state for each value of energy or no degeneracy in the system, the number of micro-states $\Omega$ is proportional to probability distribution of energy in the system.
Therefore, the number of micro-states $\Omega$ will depend on system energy which comes from each micro-state of the system. Mathematically, $\Omega$ can be written as $\Omega(E)$ where $E$ is system energy.

After entropy of the entire system is obtained, Helmholtz free-energy, $F$, and entropy, $S$, are related by definition
\begin{equation} 
F = U- TS,  
\label{eq:helmholtz} 
\end{equation}
where $U$ is internal energy and $T$ is temperature. Entropy elasticity assumes that the change of Helmholtz free energy is mostly due to the change of entropy. Therefore, internal energy can be approximately viewed as a constant. Its contribution to force and stretch is nearly zero. By ignoring the effect of $U$,  Eq.(\ref{eq:helmholtz}) is changed to 
\begin{equation} 
F = - TS.  
\label{eq:helmholtz_2} 
\end{equation}
Combining Eq.(\ref{eq:boltzmann}) and Eq.(\ref{eq:helmholtz_2}), we obtained
\begin{equation} 
F = - k_BT ln \Omega.  
\label{eq:helmholtz_3} 
\end{equation}
Helmholtz free energy is defined as free energy with constant temperature and fixed volume. For most solids, these constraints are approximately true \citep{zhao_theory_2012}. So strain energy density in solids defined as $W$ is considered to be the same as Helmholtz free energy.  
With strain energy density given as Helmholtz free energy, stress can be obtained by taking derivative of strain energy density with respect to strain or stretch. If strain energy density is based on Gaussian distribution of molecular chains, Neo-Hookean hyperleastic model can be obtained. If strain energy density is based on Langevin chains, 3 chain model, 4 chain model, and 8 chain model can be obtained. Gent model can be shown to have a close relationship with 8 chain model \citep{yang_note_2018}.  8 chain model or Arruda-Boyce model \citep{arruda_three-dimensional_1993} is widely used as a nonlinear spring for many large deformation viscoelastic models \citep{wang_mechanics_2015, jiang_physically-based_2023, uchida_viscoelastic-viscoplastic_2022}. Recently, many researchers work on rubber elasticity by other non-Gaussian chain models \citep{mezzasalma_rubber_2022, vernerey_statistically-based_2017}. Their mathematical structures are very complicate. But their ideas are still based on entropy elasticity framework. To fulfill some applications or needs,  many other phenomenological hyperelastic models such as Ogden's hyperelastic model \citep{holzapfel_nonlinear_2002} or many $I_1$ based hyperelastic models \citep{lopez-pamies_new_2010} were also built by assuming some form of strain energy density function to fit experimental data. No statistical probability distributions of microscopic molecular chains are considered in these models.
\section{Deep Into Entropy Elasticity}
\label{s:entropy} \noindent

Although entropy elasticity gains lots of success in explaining mechanical behavior of rubbers, many concepts used in entropy elasticity need a second thought.
In entropy elasticity, Helmholtz free energy is related to the number of micro-states by Eq.(\ref{eq:helmholtz_3}). As we know, the number of micro-states is very hard to calculate. For example, given a molecular chain which has 100 links, each link can elongate or contract which is assumed to have 2 micro-states. Hence, there are $2^{100} = 1.267E30$ micro-states.  But this molecular chain has only 101 macro-states. One marco-state can be elongation of all 100 links. Another marco-state could be contraction of 1 link and elongation of the rest 99 links.  Each macro-state is related to different number of micro-states. It is almost impossible to find all micro-states. But it is possible to find the probability of each macro-state. If we define the number of micro-states of a macro-state $i$ to all number of micro-states as probability $P_i$, entropy of the system  can be given as Gibbs entropy equation
\begin{equation} 
S = - k_B \sum_i P_i lnP_i.  
\label{eq:gibbs_entropy} 
\end{equation}  
By using Gibbs entropy equation, Helmholtz free energy is given as
\begin{equation} 
F = - k_B T ln Z, 
\label{eq:Helm_entropy} 
\end{equation}  
where $Z$ is partition function given as $Z = \sum_i \mathbf{e}^{-E_i/{k_BT}}$. $E_i$ is the energy of each macro-state.
So we have two Helmholtz free energy equations, e.g., Eq. (\ref{eq:helmholtz_3}) and Eq. (\ref{eq:Helm_entropy}). One is based on total micro-states and the other one is based on partition function. 
In entropy elasticity, Gaussian probability function of the end to end length vector of a molecular chain is found. Should it be viewed as the probability distribution of each micro-state to total micro-states? Or should it be viewed as the probability distribution of the number of micro-states of a macro-state $i$ to all number of micro-states, e.g., $P_i$? This is not clear. Different consideration will lead to different ways to calculation of Helmholtz free energy. Since Helmholtz free energy is used as strain energy density for hyperelastic materials, different stress-strain relationship will be got. When we solve conservation laws with different constitutive models, it will lead to different results \citep{yu_numerical_2010, chen_simulations_2011, yang_viscoelasticity_2013}.

\begin{figure}
\includegraphics[scale = 0.3]{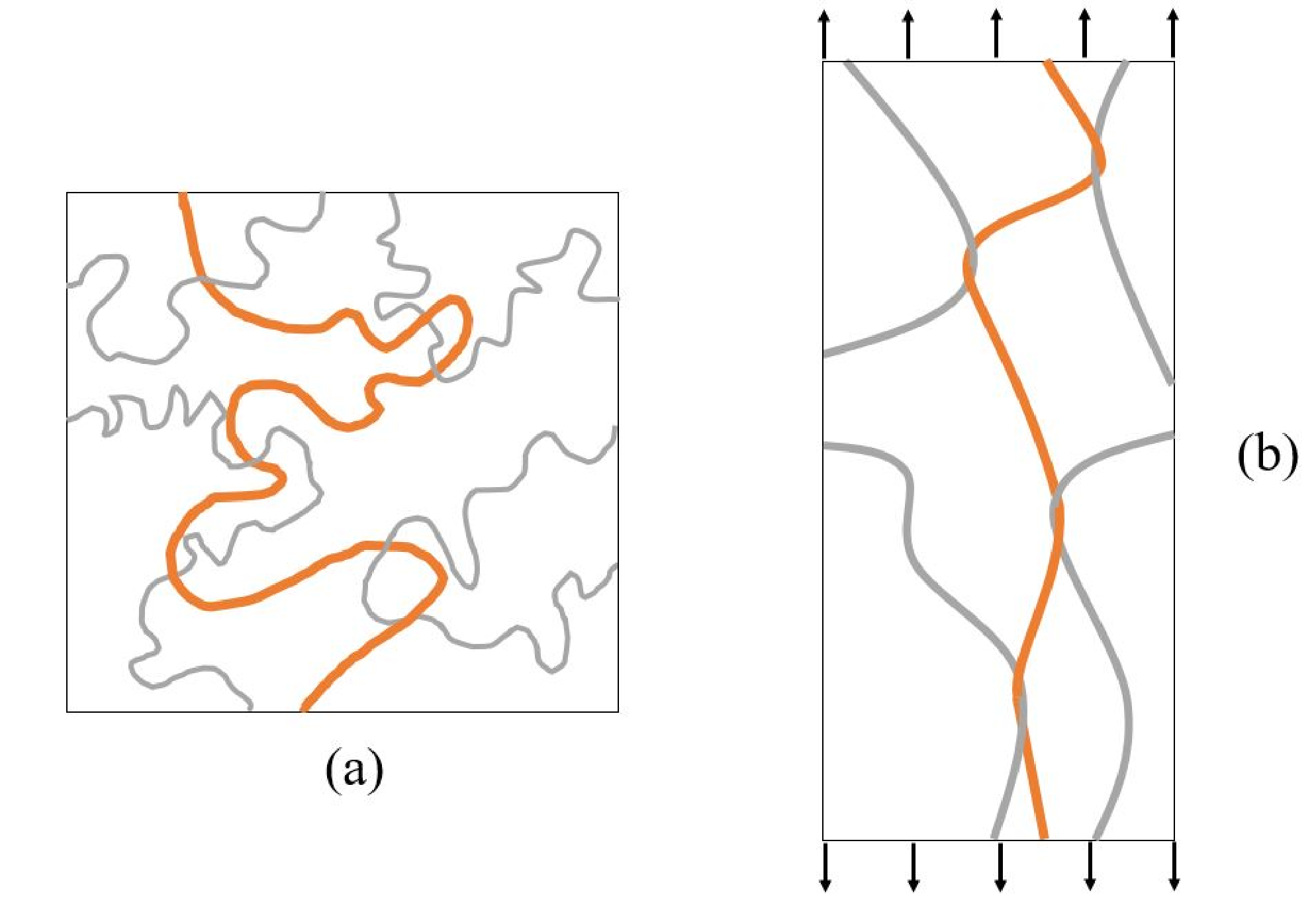}
\caption{
Schematic illustration of a molecular chain (orange color) in a relax state (a) and in a stretched state (b).
}
\label{f:Chain3}
\end{figure}

The second concern about entropy elasticity is that its firstly dealing with only one molecular chain by probability distribution. Then an affine or non-affine \citep{miehe_micro-macro_2004} or phantom assumption \citep{akagi_examination_2011} is made to extend to an entire molecular chain system.  In entropy elasticity, it assumes that a single molecular chain can have more configurations in relax states than after stretched. Since it assumes that these configurations are proportional to micro-states, micro-states and entropy will be deduced after stretched. If a single molecular chain is fully stretched to its contour length, it will have only one configuration. Therefore, entropy is the smallest. This thought is mostly based on freely jointed chain assumption. In freely jointed chains, each link can freely rotate and is rigid. No internal energy change happens when a freely jointed chain changes its configurations. 

However, in real polymer materials, polymer chains are not freely jointed. Rotation of each bond is associated with energy redistribution \citep{mulderrig_statistical_2022}.  Bonds between monomers are not rigid. Bond length is not constant. It can elongate or contract under an external force. Bond length variation will be related to energy change of the entire molecular chain.

Because bond number of a molecule chain is the same before stretched and after stretched, micro-states of a molecular chain should remain the same even a molecular chain has more configurations in a relaxed state. 
Since these micro-states can be empty or occupied by electrons, the difference of a molecular chain before stretched and after stretched is the probability redistribution of electron occupancy in these micro-states. If more electrons occupy in higher energy micro-states, entire energy of the system will be higher. That is, after stretched by an external force, more high energy micro-states are occupied by electrons in each bond of molecular chains. 
Therefore, more configurations of a molecular chain do not necessarily mean more micro-states.  Entropy is not necessarily decreasing due to stretched chains.

In entropy elasticity, since entropy is related to micro-states by Boltzmann's entropy equation, internal energy is completely separated from entropy and ignored. This is probably true for freely joint rigid chain. In fact, for real chain, those electron filled micro-states will have internal energy. Those empty micro-states will have no internal energy.  Because entropy is related to micro-states by Boltzmann entropy equation, entropy should be a function of internal energy \citep{mao_rupture_2017}. 

In entropy elasticity, stiffing effect in stress-strain curve is considered as a molecular chain reaching its contour length or its full length. However,    
even under large tensile force, a molecular chain can hardly reach its contour length. This is because that there are many randomly distribution chains instead of one molecular chain inside polymers. The stiffing effect in stress-strain curve of rubber materials can be viewed as more entanglements formed after being stretched, See Fig.(\ref{f:Chain3}). Before stretched, each molecular chain can hardly feel the exist of other molecular chains. After putting tensile force on it, molecular chains will move like snakes or reptations \citep{yang_mathematical_2019}. They will start to have interactions with other molecular chains. Many entangled forces will be built up after stretched. These gradually built-up entangled forces are the cause of stiffing effect which is shown in rubber force-stretch curve. It can be physically considered as more defects along a molecular chain. Or some parts of a molecular chain get out of its reptation tube if a size of reptation tube is assumed \citep{li_predictive_2012}. If force continues to increase, eventually the molecular chain will break. Fracture will happen \citep{kim_fracture_2021}.

Another evidence that makes people to believe that entropy plays a role in rubber elasticity is that rubbers become stiffer while metals become softer after heating them up \citep{yoshikawa_negative_2021}. Flory called this force as the force of retraction. When we observed an abnormal phenomena, there can be many explanations. Entropy can be one of them. But we think it is due to fluctuation of molecular chains instead of entropy change. Unlike crystalline metals, molecular chains are very long with random distribution. With temperature rising, each molecular chain can easily feel the other molecular chains. More temporary bonds or dynamical bonds in transient networks \citep{katashima_experimental_2022, vernerey_statistically-based_2017} can be formed due to large amplitude vibrations  of molecular chains in high temperature. Therefore, bond density increases with temperature. This will increase Young's modulus of rubbers. But this change is small and temporary. The overall change of Young's modulus from glassy zone to rubbery zone is decreasing \citep{yang_viscoelasticity_2021}.

Entropy elasticity is originally built by considering static loading with constant temperature and fixed pressure. In fact, the contraints of Helmholtz free energy are constant temperature and fixed volume.
Hence, it's hard for existed entropy elasticity to extend to  temperature and hydrostatic pressure domain \citep{katashima_decoupling_2022}. It is also not easy to extend to time domain. This is because entropy elasticity is built on equilibrium thermodynamics \citep{zhang_statistical-chain-based_2023}. First law of thermodynamics tell us that energy will be conserved. but didn't tell us how fast and how slow the energy flows. To extend statistical elasticity to time, temperature and hydrostatic pressure dependent area, we believe that it will rely on how to describe bond deformations of molecular chains \citep{buche_freely_2022}. To consider this, statistical description of covalent electrons should be used since each bond is formed by covalent electrons.  Borrowed from statistical description of gas molecules, current entropy elasticity theory is not appropriate to understand these bond movements. The big difference between gas molecules and covalent electrons in solids is that the distance between two gas molecules is very large comparing with the distance between two electrons. Physically, the distance between two gas molecules is very larger than "thermal" de Broglie wavelength. The moving path of one gas molecule can be distinguished from that of another gas molecular. They are treated as distinguishable particles. The distance between two covalent electrons is comparable or smaller than "thermal" de Broglie wavelength. They are fermions and obey the Pauli exclusion principle. They are treated as indistinguishable particles.

For distinguishable particles such as gas molecules, classical statistical method such as Boltzmann's entropy equation and Maxwell-Boltzmann distribution are good candidates. For solids like rubbers, molecular chains are glued by covalent electrons which are indistinguishable particles, quantum statistical method such as Fermi-Dirac statistics need be used. 

In the next section, we will show how we can go from Maxwell-Boltzmann distribution to Fermi-Dirac distribution based on a molecular chain structure.

\section{From Gaussian distribution to Maxwell-Boltzmann distribution to Fermi-Dirac distribution}
\label{s:new} \noindent
Firstly let us exam Eq.(\ref{eq:Gauss_1D}) in detail.
One dimensional Gaussian distribution for end to end distance of a molecular chain, e.g., Eq.(\ref{eq:Gauss_1D}) can be written as
\begin{equation} 
G(x, N) = \frac{\mathbf{e}^{-3x^2/2Nb^2}}{(\,\frac{3}{2 \pi Nb^2} )\,^{1/2}}.   
\label{eq:Gauss_1D_a} 
\end{equation}
or
\begin{equation} 
G(x, N) = \frac{\mathbf{e}^{-kx^2/2}}{L}   
\label{eq:Gauss_1D_b} 
\end{equation}
if let $k=3/Nb^2$ and $L=(\,\frac{3}{2 \pi Nb^2} )\,^{1/2}$. Eq.(\ref{eq:Gauss_1D_b}) can be further written as
\begin{equation} 
G(x, N) = P(\textrm{microstate in x direction}) = \frac{\mathbf{e}^{-E_x}}{\sum_i \mathbf{e}^{-E_i}}   
\label{eq:Gauss_1D_c} 
\end{equation}
if let energy $E_x = kx^2/2$ and $L = \sum_i \mathbf{e}^{-E_i}$. 

However, from classical thermodynamics, we know probability of micro-state $r$ can be written as
\begin{equation} 
P(\textrm{microstate r}) = \frac{\mathbf{e}^{-E_r/k_BT}}{\sum_i \mathbf{e}^{-E_i/k_BT}}   
\label{eq:Gauss_1D_d} 
\end{equation}
where $\sum_i \mathbf{e}^{-E_i/k_BT}$ is partition function.

If comparing Eq.(\ref{eq:Gauss_1D_c}) with Eq.(\ref{eq:Gauss_1D_d}) , we know that $k_BT$ is missing in one dimensional Gaussian distribution for end to end distance of a molecular chain, e.g., Eq.(\ref{eq:Gauss_1D}). 
Should we add $k_BT$ into Gaussian distribution, e.g., Eq.(\ref{eq:Gauss_1D})? The answer is yes. In the following section, we will illustrate the reason why $k_BT$ need be added.

\begin{figure}
\includegraphics[scale = 0.25]{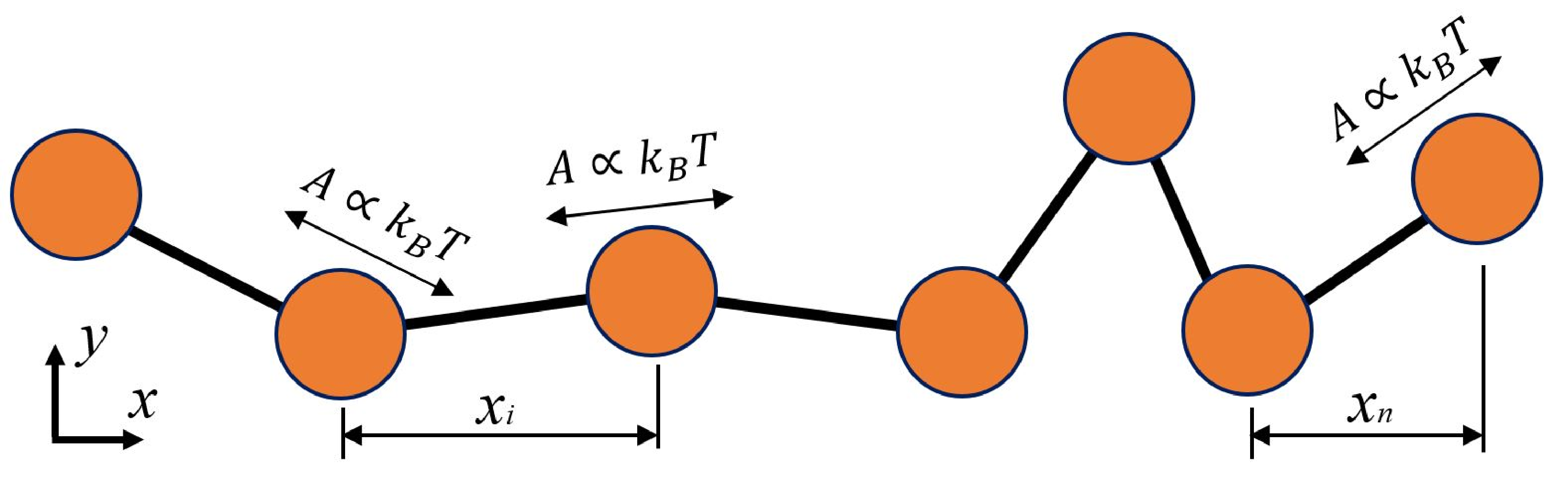}
\caption{
Schematic illustration of vibration of monomers due to temperature change. Vibration amplitude $A$ is proportional to $k_BT$.
}

\label{f:f1}
\end{figure} 

If each bond segment length vector of a molecular chain is projected to x-axis, segment length vector in x-direction, e.g., $X_i$, can be considered as a random variable, the total chain length vector in $x$ direction can be the summation of $X_i$, e.g., $S_n = X_1 +...+X_n$. 
For each random variable $X_i$, it measures the distance between two monomers in x-direction, see Fig.(\ref{f:f1}). For all random variables $X_i$, let's assume that they are independent and have the same mean $\nu$ and the same standard deviation $\sigma$. When temperature is increasing, monomer vibration amplitude will increase. Assuming that all monomers vibrate in their equilibrium positions, average distance between two monomers will remain the same. But variation of the distance between two monomers will increase. In other words, mean $\nu$ of $X_i$ remains the same while the standard deviation $\sigma$ of all $X_i$ will increase with temperature.  Even though we don't know exact probability distribution of $X_i$, 
central limit theorem (\ref{eq:CLT2}) tells us that total chain length vector will become Gaussian distribution which can be written as
\begin{equation} 
Z_n = \frac{S_n - n \nu}{\sqrt{n} \sigma}.   
\label{eq:CLT1} 
\end{equation}
So the random variable $S_n$ which is used to measure the end-to-end length vector of a molecular chain in x-direction is a normal distribution with mean $n \nu$ and standard deviation $\sqrt{n} \sigma$. Since $\sigma$ will depend on temperature $T$, we can select that  $\sigma^2 \varpropto k_BT$. 

In real polymer materials, different molecular chains can have different bond numbers. As long  as unit number $n$ of a molecular chain is larger than a critic number, all primary molecular chains will converge to normal distribution. This critic number is usually very small if probability distribution function of each bond length is symmetric. If primary molecular chains are cross-linked, central limit theorem can also be used. As long as the unit number of cross-linked molecular chain is large enough, summation of random variables following each path of cross-linked molecular chain could be viewed as normal distribution.
 
Let $\bar{\nu}$ be the average of all means of molecular chains. Let $\bar{\sigma}$ be the average of all standard deviations of molecular chains by defining $\bar{\sigma}^2=k_BT/k_0$. $k_0$ is a proportional constant which is not necessarily equal to $k=3/Nb^2$. Then Gaussian distribution function of $S_n$ can be written as
\begin{equation}  
G(x, N) = \frac{\sqrt{k_0}}{\sqrt{2\pi k_BT}}\mathbf{e}^{-\frac{k_0}{2k_BT}\left(x-\bar{\nu} \right)^2}.  
\label{eq:Gauss_1D_e} 
\end{equation}
By using equal partition theorem and let $\bar{\nu} \rightarrow 0$, Eq.(\ref{eq:Gauss_1D_e}) becomes Maxwell-Boltzmann distribution in Cartesian coordinate system, e.g.,
\begin{equation}  
G(x, N) = \sqrt{\frac{m}{2\pi k_BT}}\mathbf{e}^{-\frac{m{v_x}^2}{2k_BT}},  
\label{eq:Gauss_1D_f} 
\end{equation}
where $m$ is mass of gas molecules and $v_x$ is velocity of gas molecules in x-direction.
Velocity distribution of gas molecules can be found by Maxwell-Boltzmann distribution in polar coordinate system.
Maxwell-Boltzmann distribution can be viewed as temperature dependent Gaussian distribution.
Here, we use Maxwell-Boltzmann distribution to describe probability distribution of the end-to-end length vector of a molecular chain with an external force. When an external force is applied to the molecular chain,  electrons will move from low energy states to high energy states. Probability distribution of the end-to-end length vector of a molecular chain can be described by Maxwell-Boltzmann distribution. But energy storage inside the deformed molecular chain is related to electron occupancy in each quantum micro-states. The probability to find electron occupancy in each quantum micro-state need be described by Fermi-Dirac function. Once we know the end-to-end length vector of a molecular chain, energy storage due to electron occupancy in each quantum micro-state can be found. 
In another word, we can derive Fermi-Dirac function by using Maxwell-Boltzmann distribution, e.g., Eq.(\ref{eq:Gauss_1D_e}).

When an external force is applied to polymer materials,  this force will do work on polymer materials. Some of the work will become strain energy absorbed by molecular chains. The configuration of molecular chains will get changed.  Because of the randomness of molecular chains, we don't know how each bond changes its rotation angle and bond length. We only know the probability distribution of the end to end length vector of each polymer chain obeys Maxwell-Boltzmann distribution. 

Firstly let's consider a signal transmission problem. If an unknown signal (probable inputs are assumed to be 1/2 and -1/2) is sent to polymer materials, this signal will go through these molecular chains. Measured output signal will be contaminated by those Maxwell-Boltzmann distributed molecular chains, which can be viewed as a random noise. Output signal is a combination of random noise and input signal and becomes a continuous random signal. If we can measure the output continuous random signal, probability to find which input signal is sent with known measured output signal is given by mixed Bayes rule.
\begin{figure}
\centering \includegraphics[scale = 0.3]{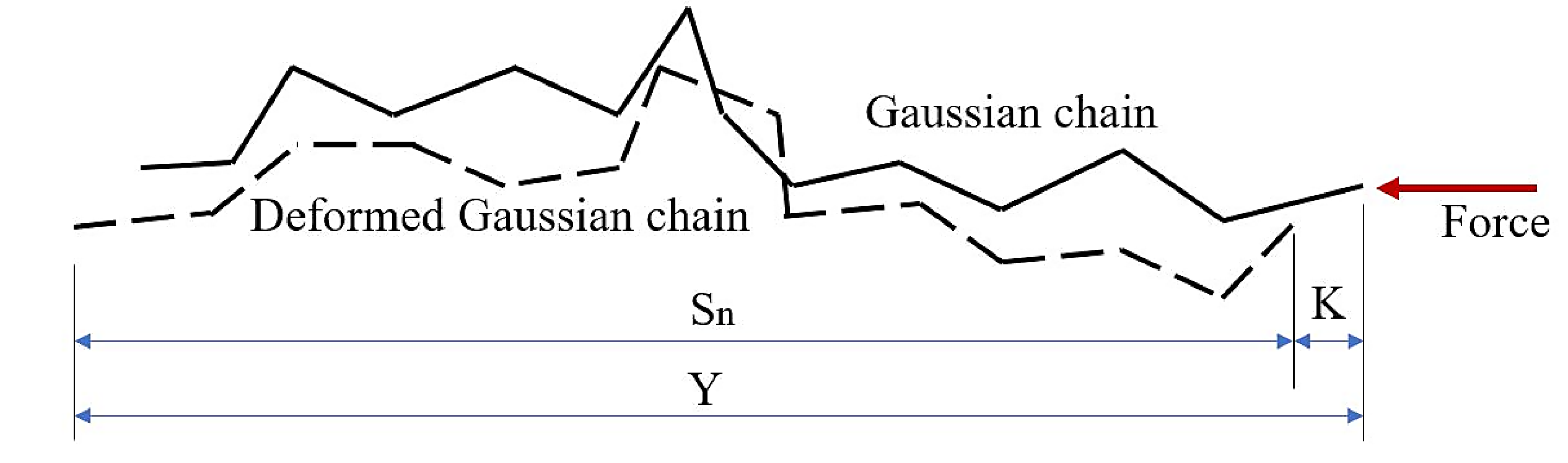}
\caption{
Representation of a Gaussian molecular chain in two dimensions under a random external force. $K$, $S_n$ and $Y$ are considered as random variables.
}

\label{f:Chain1}
\end{figure}

Analogy to this signal problem, consider the length change of a molecular chain under an input force. 
Let's assume that an unknown force is applied at the starting point of the chain, see Fig.(\ref{f:Chain1}). The displacement at the starting point due to this unknown force is defined as a discrete random variable $K$. 
 Without losing generality, let probability to be 0.5 if this end moves to the left by $-0.5$  and probability to be 0.5 if it moves to the right by $0.5$. That is, the discrete random variable $K$ is defined as
\begin{equation}
 P_K(k) = \begin{cases}
0.5, \text{if $k=0.5$} \\
0.5, \text{if $k=-0.5$}
  \end{cases}
  \label{eq:discrete_ex1}
\end{equation}

Recall that $S_n$ is a continuous random variable of the end to end length vector of a  molecular chain with an external force, see Fig.(\ref{f:Chain1}). 
After imposing a force on the starting point of a molecular chain, we want to locate the ending point of the molecular chain. 
Let's define a continuous random variable $Y$  which measures the total length from the ending point of this deformed molecular chain to the starting point of this initial undeformed chain, See Fig.(\ref{f:Chain1}). So mathematically $Y = K + S_n$. That is,  total length is the combination of input displacement and end to end length of a molecular chain. 

Firstly, let's consider two extreme cases when molecular chains are extremely stiff or extremely soft. If a molecular chain is super stiff, variation of each bond length will be close to $0$. Temperature dependent Gaussian distribution (Maxwell-Boltzmann distribution) of a molecular chain $S_n$ will collapse to delta function with constant mean and variance is nearly $0$. For example, assuming an input displacement $K$ is $1/2$ and the length of a molecular chain $S_n$ is $10$, output length $Y$ will be $10.5$. In other words, the probability of finding input displacement is $1/2$ given measured output length $10.5$ is $1$. It also says, the probability that available micro-states are occupied by electrons in this super stiff molecular chain is $1$.

If a molecular chain is super soft, variance of temperature dependent Gaussian distribution will approach infinity. Probability distribution of the end to end distance vector of a molecular chain can be viewed as uniform distribution from negative infinity to positive infinity. Hence, it is hard to determine mean value of $S_n$. Assuming input displacement is $1/2$ and the length of a molecular chain is 10, output length can not be determined at all because the variance of this super soft chain is infinity. In other words, the probability of finding input displacement is $1/2$ given any measured output length is $0$. This gives us an idea that available micro-states are all empty in this super soft molecular chain. Or the probability that available micro-states are occupied by electrons is $0$.

Between those two extreme cases when a molecular chain obeys temperature dependent Gaussian distribution, probability of available micro-states occupied by electrons is between $0$ and $1$. 

Therefore, probability to find input displacement, e.g., $K$, imposed on a molecular chain with given measured total length of a deformed molecular chain $Y$ will be the probability that available micro-states are occupied by electrons.

Even though we consider only one molecular chain configuration here, Eq. (\ref{eq:Gauss_1D_e}) is applied to all molecular chains since average mean and average variance are used. 

By using mixed Bayes rule, the probability of finding input displacement $k$ given any measured output length $y$ is given as posterior probability distribution, e.g., $P_{K|Y}(k|y)$. It also tells us the probability that available micro-states are occupied by electrons. To obtain the posterior probability distribution, e.g., $P_{K|Y}(k|y)$, prior probability distribution $f_{Y|K}(y|k)$ and $f_Y(y)$ need be calculated firstly. Since the random variable of measured output length $Y = K + S_n$, prior probability distribution $f_{Y|K}(y|k)$ is a conditional Maxwell-Boltzmann distribution. That is, if we know the exact displacement at the starting point of a molecular chain, the probability to find the other end of a molecular chain is given by conditional Maxwell-Boltzmann distribution $f_{Y|K}(y|k)$. $f_Y(y)$ is the probability to find the other end of a molecular chain regardless of input displacement.  
 
A brief review of mixed Bayes rule is given in the appendix. Bayesian inference has also been used to evaluate parameters of constitutive models and polymer properties \citep{thomas_bayesian_2022, long_experiments_2022}.

To proceed, the conditional Maxwell-Boltzmann distribution $f_{Y|K}(y|k)$ can be written as
\begin{equation}   
f_{Y|K}(y|k) \: = \: \sqrt{\frac{k_0}{2\pi k_BT}} e^{-\frac{k_0}{2k_BT}(y-\bar{\nu}-k)^2}.
\label{eq:con_Gaussian}
\end{equation}
In order to use mixed Bayes rule, $f_Y(y)$ is calculated by
\begin{equation}   
f_Y(y) \: = \: \frac{1}{2}\sqrt{\frac{k_0}{2\pi k_BT}} e^{-\frac{k_0}{2k_BT}(y-\bar{\nu}-\frac{1}{2})^2} + \frac{1}{2}\sqrt{\frac{k_0}{2\pi k_BT}} e^{-\frac{k_0}{2k_BT}(y-\bar{\nu}+\frac{1}{2})^2}.
\label{eq:total_prob}
\end{equation}
Substituting Eq.(\ref{eq:discrete_ex1}), Eq.(\ref{eq:con_Gaussian}) and Eq. (\ref{eq:total_prob}) into Eq.(\ref{eq:Bayes}), we can get the probability of input distance is $-1/2$ if given measured total distance $y$ as
\begin{equation}   
P_{K|Y}(-\frac{1}{2}|y) \: = \: \frac{1}{e^{\frac{k_0(y-\bar{\nu})}{k_BT}}+1}.
\label{eq:Bayes_input1}
\end{equation}
By letting $\varepsilon=k_0y$ and $\mu=k_0\bar{\nu}$, Eq.(\ref{eq:Bayes_input1}) is recast as
\begin{equation}   
P_{K|Y}(-\frac{1}{2}|y) \: = \: \frac{1}{e^{\frac{\varepsilon-\mu}{k_BT}}+1},
\label{eq:Bayes_input2}
\end{equation}
which is Fermi-Dirac distribution function. Fermi Dirac distribution is the probability that an available energy micro-state at energy level $\varepsilon$ is occupied by an electron.
$\mu$ is called chemical potential or Fermi energy. 
Fermi-Dirac distribution can be used to calculate energy storage in each bond of a molecular chain if we know the density of micro-states.
Similarly, the probability of input distance is $1/2$ if given measured total distance $y$  is given as
\begin{equation}   
P_{K|Y}(\frac{1}{2}|y) \: = \: \frac{1}{e^{-\frac{\varepsilon-\mu}{k_BT}}+1},
\label{eq:Bayes_input3}
\end{equation}
which is the probability of finding a hole at an available energy micro-state.
The probability of finding an electron in an available micro-state is one minus the probability of finding a hole in an available micro-state, e.g.,
\begin{equation}   
\frac{1}{e^{\frac{\varepsilon-\mu}{k_BT}}+1}  \: = \: 1- \frac{1}{e^{-\frac{\varepsilon-\mu}{k_BT}}+1}.
\label{eq:Bayes_input4}
\end{equation}

In polymer materials, cross-links and entanglements are co-existed. As we discussed, cross-links are bonds formed by covalent electrons, which can be modeled by Fermi-Dirac distribution.  Entanglements need be considered separately. They are weak forces compared to forces caused by covalent bonds. Entanglements can be viewed as a transient network. Density and velocity of entanglements can be easily changed by an external force. A possible treatment of entanglements is considering them as defects along molecular chains \citep{yang_mathematical_2019}. In another word, defect density and defect velocity will change under an external force. Idea of this treatment is consistent with tube and reptation models \citep{li_predictive_2012}. It will be treated as a defect if any part of a molecular chain gets out of the assumed tube under an external force. The mathematical structure of Fermi-Dirac distribution is also much easier and cleaner than the slip-link model introduced by Edwards and Vilgis \citep{davidson_nonaffine_2013, edwards_effect_1986}.   

Mathematically, Fermi-Dirac distribution function is logistic function. Logistic function can be generalized to include time, temperature, hydrostatic pressure effects. By putting generalized Fermi-Dirac distribution into our stress-strain equation, many mechanical behavior problems of polymers can be easily understood. For example, shear band and necking formation is localized plastic deformation if defect effect is larger than bonding effect \citep{yang_revisit_2020}. Ultrasonic vibration is considered to be an inverse process of shear band generation \citep{yang_digest_2023}. Knee injury can be prevented by erasing physical aging which can be understood by time dependent version of generalized Fermi-Dirac distribution \citep{yang_theoretical_2020}. Generalized logistic function has been used to model linear viscoelasticity \citep{yang_viscoelasticity_2021}. Time, temperature, and hydrostatic pressure dependent viscosity of polymers can be understood by generalized Fermi-Dirac distribution. 
Aided by defect concepts in molecular chains, modeling nonlinear viscoplasiticty of polymers is straightforward \citep{yang_mathematical_2019}. Model prediction is validated by many experiments shown in our previous papers\citep{yang_mathematical_2019,yang_viscoelasticity_2021}.

\section{Conclusion} 
\label{s:conclusion}
In this article, we showed how to derive from temperature independent Gaussian distribution to temperature dependent Gaussian distribution or Maxwell-Boltzmann distribution. Then Fermi-Dirac distribution is derived from Maxwell-Boltzmann distribution based on a deformed molecular chain under an external force. 
Since the distance between gas molecules is very larger than "thermal" de Broglie wavelength, Maxwell-Boltzmann distribution should be used for describing the energy distribution for gas molecules. On the other hand, Fermi-Dirac distribution is needed to capture energy distribution for covalent bonds in polymer molecular chains. Coupled with defects or entanglements along molecular chains, generalized Fermi-Dirac distribution can be used to explain most if not all polymer mechanical problems. Since human tissue is made of molecular chains of different bond stiffness, this method will be extended to explain intra-cranial brain tissue dynamics in the next paper.

\section*{Data Availability}
Data sharing not applicable - no new data generated.
\section*{Reference}
\bibliographystyle{elsarticle-num}
\bibliography{shang_dft}

\begin{thebibliography}{10}
\expandafter\ifx\csname url\endcsname\relax
  \def\url#1{\texttt{#1}}\fi
\expandafter\ifx\csname urlprefix\endcsname\relax\def\urlprefix{URL }\fi
\expandafter\ifx\csname href\endcsname\relax
  \def\href#1#2{#2} \def\path#1{#1}\fi

\bibitem{yang_viscoelasticity_2021}
L.~Yang, L.~Yang, R.~L. Lowe, A viscoelasticity model for polymers: {Time},
  temperature, and hydrostatic pressure dependent {Young}'s modulus and
  {Poisson}'s ratio across transition temperatures and pressures, Mechanics of
  Materials 157 (2021) 103839.

\bibitem{yang_mathematical_2019}
L.~Yang, A {Mathematical} {Model} for {Amorphous} {Polymers} {Based} on
  {Concepts} of {Reptation} {Theory}, Polymer Engineering \& Science 59~(11)
  (2019) 2335--2346.

\bibitem{yang_revisit_2020}
L.~Yang, L.~Yang, Revisit initiation of localized plastic deformation: {Shear}
  band \& necking, Extreme Mechanics Letters 40 (2020) 100914.

\bibitem{yang_digest_2023}
L.~Yang, L.~Yang, Digest {Ultrasonic} {Welding} {I}: {Localized} {Heating} and
  {Fuse} {Bonding}, arXiv preprint arXiv:2301.05810 (2023) 1--13.

\bibitem{zhan_new_2023}
L.~Zhan, S.~Wang, S.~Qu, P.~Steinmann, R.~Xiao, A new micro–macro transition
  for hyperelastic materials, Journal of the Mechanics and Physics of Solids
  171 (2023) 105156.

\bibitem{li_effects_2021}
C.~Li, Z.~Wang, Y.~Wang, Q.~He, R.~Long, S.~Cai, Effects of network structures
  on the fracture of hydrogel, Extreme Mechanics Letters 49 (2021) 101495.

\bibitem{wang_mechanics_2015}
Q.~Wang, G.~R. Gossweiler, S.~L. Craig, X.~Zhao, Mechanics of mechanochemically
  responsive elastomers, Journal of the Mechanics and Physics of Solids 82
  (2015) 320--344.

\bibitem{akagi_examination_2011}
Y.~Akagi, T.~Katashima, Y.~Katsumoto, K.~Fujii, T.~Matsunaga, U.-i. Chung,
  M.~Shibayama, T.~Sakai, Examination of the {Theories} of {Rubber}
  {Elasticity} {Using} an {Ideal} {Polymer} {Network}, Macromolecules 44~(14)
  (2011) 5817--5821.

\bibitem{buche_chain_2021}
M.~R. Buche, M.~N. Silberstein, Chain breaking in the statistical mechanical
  constitutive theory of polymer networks, Journal of the Mechanics and Physics
  of Solids 156 (2021) 104593.

\bibitem{zhang_two-component_2022}
F.~Zhang, S.~Cui, A two-component statistical model for natural rubber, Polymer
  242 (2022) 124462.

\bibitem{zhang_single-chain_2022}
F.~Zhang, Z.~Gong, W.~Cai, H.-j. Qian, Z.-y. Lu, S.~Cui, Single-chain mechanics
  of cis-1,4-polyisoprene and polysulfide, Polymer 240 (2022) 124473.

\bibitem{flory_principles_1953}
P.~J. Flory, Principles of {Polymer} {Chemistry}, Cornell University Press,
  1953.

\bibitem{darabi_amended_2023}
E.~Darabi, M.~Hillgärtner, M.~Itskov, An amended approximation of the
  non-{Gaussian} probability distribution function, Mathematics and Mechanics
  of Solids 28~(2) (2023) 521--532.

\bibitem{tanaka_viscoelastic_1992}
F.~Tanaka, S.~F. Edwards, Viscoelastic properties of physically crosslinked
  networks. 1. {Transient} network theory, Macromolecules 25~(5) (1992)
  1516--1523.

\bibitem{vernerey_statistical_2018}
F.~J. Vernerey, R.~Brighenti, R.~Long, T.~Shen, Statistical damage mechanics of
  polymer networks, Macromolecules 51~(17) (2018) 6609--6622.

\bibitem{mahnken_statistically_2022}
R.~Mahnken, J.~Mirzapour, A statistically based strain energy function for
  polymer chains in rubber elasticity, Archive of Applied Mechanics 92~(11)
  (2022) 3295--3323.

\bibitem{jiang_strain_2022}
Y.~Jiang, L.~Li, Y.~Hu, Strain gradient elasticity theory of polymer networks,
  Acta Mechanica 233~(8) (2022) 3213--3231.

\bibitem{zhao_theory_2012}
X.~Zhao, A theory for large deformation and damage of interpenetrating polymer
  networks, Journal of the Mechanics and Physics of Solids 60~(2) (2012)
  319--332.

\bibitem{yang_note_2018}
L.~Yang, L.~Yang, Note on {Gent}'s hyperelastic model, Rubber Chemistry and
  Technology 91~(1) (2018) 296--301.

\bibitem{arruda_three-dimensional_1993}
E.~M. Arruda, M.~C. Boyce, A three-dimensional constitutive model for the large
  stretch behavior of rubber elastic materials, Journal of the Mechanics and
  Physics of Solids 41~(2) (1993) 389--412.

\bibitem{jiang_physically-based_2023}
Y.~Jiang, L.~Li, Y.~Hu, A physically-based nonlocal strain gradient theory for
  crosslinked polymers, International Journal of Mechanical Sciences 245 (2023)
  108094.

\bibitem{uchida_viscoelastic-viscoplastic_2022}
M.~Uchida, K.~Kamimura, T.~Yoshida, Y.~Kaneko, Viscoelastic-viscoplastic
  modeling of epoxy based on transient network theory, International Journal of
  Plasticity 153 (2022) 103262.

\bibitem{mezzasalma_rubber_2022}
S.~A. Mezzasalma, M.~Abrami, G.~Grassi, M.~Grassi, Rubber elasticity of polymer
  networks in explicitly non-{Gaussian} states. {Statistical} mechanics and
  {LF}-{NMR} inquiry in hydrogel systems, International Journal of Engineering
  Science 176 (2022) 103676.

\bibitem{vernerey_statistically-based_2017}
F.~J. Vernerey, R.~Long, R.~Brighenti, A statistically-based continuum theory
  for polymers with transient networks, Journal of the Mechanics and Physics of
  Solids 107 (2017) 1--20.

\bibitem{holzapfel_nonlinear_2002}
G.~A. Holzapfel, Nonlinear {Solid} {Mechanics}: {A} {Continuum} {Approach} for
  {Engineering} {Science}, Meccanica 37~(4) (2002) 489--490.

\bibitem{lopez-pamies_new_2010}
O.~Lopez-Pamies, A new {I1}-based hyperelastic model for rubber elastic
  materials, Comptes Rendus Mécanique 338~(1) (2010) 3--11.

\bibitem{yu_numerical_2010}
S.-T.~J. Yu, L.~Yang, R.~L. Lowe, S.~E. Bechtel, Numerical simulation of linear
  and nonlinear waves in hypoelastic solids by the {CESE} method, Wave Motion
  47~(3) (2010) 168--182.

\bibitem{chen_simulations_2011}
Y.-Y. Chen, L.~Yang, S.-T.~J. Yu, Simulations of waves in elastic solids of
  cubic symmetry by the conservation element and solution element method, Wave
  Motion 48~(1) (2011) 39--61.

\bibitem{yang_viscoelasticity_2013}
L.~Yang, Y.-Y. Chen, S.-T.~J. Yu, Viscoelasticity determined by measured wave
  absorption coefficient for modeling waves in soft tissues, Wave Motion 50~(2)
  (2013) 334--346.

\bibitem{miehe_micro-macro_2004}
C.~Miehe, S.~Göktepe, F.~Lulei, A micro-macro approach to rubber-like
  materials—{Part} {I}: the non-affine micro-sphere model of rubber
  elasticity, Journal of the Mechanics and Physics of Solids 52~(11) (2004)
  2617--2660.

\bibitem{mulderrig_statistical_2022}
J.~Mulderrig, B.~Talamini, N.~Bouklas, A statistical mechanics framework for
  polymer chain scission, based on the concepts of distorted bond potential and
  asymptotic matching (Aug. 2022).

\bibitem{mao_rupture_2017}
Y.~Mao, B.~Talamini, L.~Anand, Rupture of polymers by chain scission, Extreme
  Mechanics Letters 13 (2017) 17--24.

\bibitem{li_predictive_2012}
Y.~Li, S.~Tang, B.~C. Abberton, M.~Kröger, C.~Burkhart, B.~Jiang, G.~J.
  Papakonstantopoulos, M.~Poldneff, W.~K. Liu, A predictive multiscale
  computational framework for viscoelastic properties of linear polymers,
  Polymer 53~(25) (2012) 5935--5952.

\bibitem{kim_fracture_2021}
J.~Kim, G.~Zhang, M.~Shi, Z.~Suo, Fracture, fatigue, and friction of polymers
  in which entanglements greatly outnumber cross-links, Science 374~(6564)
  (2021) 212--216.

\bibitem{yoshikawa_negative_2021}
Y.~Yoshikawa, N.~Sakumichi, U.-i. Chung, T.~Sakai, Negative energy elasticity
  in a rubberlike gel, Physical Review X 11~(1) (2021) 011045.

\bibitem{katashima_experimental_2022}
T.~Katashima, R.~Kudo, M.~Naito, S.~Nagatoishi, K.~Miyata, U.-i. Chung,
  K.~Tsumoto, T.~Sakai, Experimental {Comparison} of {Bond} {Lifetime} and
  {Viscoelastic} {Relaxation} in {Transient} {Networks} with
  {Well}-{Controlled} {Structures}, ACS Macro Letters 11~(6) (2022) 753--759.

\bibitem{katashima_decoupling_2022}
T.~Katashima, R.~Kobayashi, S.~Ishikawa, M.~Naito, K.~Miyata, U.-i. Chung,
  T.~Sakai, Decoupling between {Translational} {Diffusion} and
  {Viscoelasticity} in {Transient} {Networks} with {Controlled} {Network}
  {Connectivity}, Gels 8~(12) (2022) 830.

\bibitem{zhang_statistical-chain-based_2023}
H.~Zhang, Y.~Hu, A statistical-chain-based theory for dynamic living polymeric
  gels with concurrent diffusion, chain remodeling reactions and deformation,
  Journal of the Mechanics and Physics of Solids 172 (2023) 105155.

\bibitem{buche_freely_2022}
M.~R. Buche, M.~N. Silberstein, S.~J. Grutzik, Freely jointed chain models with
  extensible links, Physical Review E 106~(2) (2022) 024502.

\bibitem{thomas_bayesian_2022}
A.~J. Thomas, E.~Barocio, I.~Bilionis, R.~B. Pipes, Bayesian inference of fiber
  orientation and polymer properties in short fiber-reinforced polymer
  composites, Composites Science and Technology 228 (2022) 109630.

\bibitem{long_experiments_2022}
T.~Long, S.~Shende, C.-Y. Lin, K.~Vemaganti, Experiments and hyperelastic
  modeling of porcine meniscus show heterogeneity at high strains, Biomechanics
  and Modeling in Mechanobiology 21~(6) (2022) 1641--1658.

\bibitem{davidson_nonaffine_2013}
J.~D. Davidson, N.~C. Goulbourne, A nonaffine network model for elastomers
  undergoing finite deformations, Journal of the Mechanics and Physics of
  Solids 61~(8) (2013) 1784--1797.

\bibitem{edwards_effect_1986}
S.~F. Edwards, T.~Vilgis, The effect of entanglements in rubber elasticity,
  Polymer 27~(4) (1986) 483--492.

\bibitem{yang_theoretical_2020}
L.~Yang, Theoretical and numerical analysis of anterior cruciate ligament
  injury and its prevention, Global Journals of Research in Engineering 20~(J1)
  (2020) 43--54.

\end{thebibliography}

\pagebreak
\begin{appendices}
\section{Brief review of central limit theorem and mixed Bayes rule}
\label{s:statistics} \noindent
The basic idea of central limit theorem is that the summation of many individual independent random variables will converge to the same Gaussian distribution.
Let $X_1, ..., X_n$ are $n$ identical independent distributed random variables with finite mean $\nu$ and finite variance $\sigma^2$. These $n$ random variables need not be normal distribution. $S_n$ is defined as the sum of these $n$ identical independent distributed random variables, e.g., $S_n = X_1 + ... + X_n$. A new random variable can be defined as
\begin{equation} 
Z_n = \frac{S_n - n \nu}{\sqrt{n} \sigma}.   
\label{eq:CLT1} 
\end{equation}
If $Z$ is a standard normal random variable with zero mean and unit variance, central limit theorem tell us  that $Z_n$ will be approach to $Z$ if $n$ goes to infinity. Mathematically central limit theorem is expressed as
\begin{equation}  
Z_n = \lim_{n \to \infty} P(Z_n \leq z) = P(Z \leq z).   
\label{eq:CLT2} 
\end{equation}

Next, we will review mixed Bayes rule which mixes a discrete random variable and a continuous random variable. Assume that we have a discrete random variable $K$ and a continuous random variable $Y$. For the discrete random variable $K$, we can define probability mass function as
\begin{equation} 
P_K(k) \: = \: P(K=k).  
\label{eq:discrete} 
\end{equation}
Conditional probability mass function for discrete random variable $K$ given continuous random variable $Y$  taking a value $y$ is defined as
\begin{equation} 
P_{K|Y}(k|y) \: = \: P(K=k|y \leq Y \leq y + \delta),  
\label{eq:con_discrete} 
\end{equation}  
where $\delta$ is a small increment. For the continuous random variable $Y$, probability density function is defined as
\begin{equation}
f_Y(y) \delta \: = \: P(y \leq Y \leq y+\delta).
\label{eq:continuous}
\end{equation}
Conditional probability density function for continuous random variable $Y$ given discrete random variable $K$ taking a value $k$ is defined as
\begin{equation}   
f_{Y|K}(y|k) \delta \: = \: P(y \leq Y \leq y+\delta | K=k).
\label{eq:con_continuous}
\end{equation}
So mixed Bayes rule can be written as
\begin{equation}   
P_{K|Y}(k|y) \: = \: \frac{P_K(k)f_{Y|K}(y|k)}{f_Y(y)},
\label{eq:Bayes}
\end{equation}
where 
\begin{equation*}   
f_Y(y) \: = \: \sum_{k'} p_K(k')f_{Y|K}(y|k').
\label{eq:Bayes_1}
\end{equation*}

\end{appendices}

\end{document}